\documentclass[11pt,a4paper]{article}
\pdfoutput=1
\usepackage{geometry}
\usepackage[pdftex]{graphicx}
\usepackage{color}
\usepackage{framed}
\usepackage[bottom]{footmisc}
\usepackage{amssymb}
\usepackage{latexsym}
\usepackage{amsmath}
\usepackage{enumitem}
\usepackage{bm}

\newcommand{\Gam}[2]{\Gamma\!\left(\tfrac{#1}{#2}\right)}
\usepackage[width=0.9\textwidth]{caption}
\usepackage[retainorgcmds]{IEEEtrantools}
\usepackage[bookmarks=true,pdfborder={0 0 0}]{hyperref}
\usepackage[all]{hypcap}
\title{Analytic vortex solutions on compact hyperbolic surfaces}
\author{Rafael Maldonado\footnote{{\tt R.Maldonado@damtp.cam.ac.uk}}\hspace{0.25cm}and Nicholas S.~Manton\footnote{{\tt N.S.Manton@damtp.cam.ac.uk}}\vspace{0.2cm}\\\emph{Department of 
Applied Mathematics and Theoretical Physics,}\\\emph{Wilberforce Road, Cambridge CB3 0WA, U.K.}}

\begin{document}
\maketitle
\begin{abstract}
\noindent We construct, for the first time, Abelian-Higgs vortices on certain compact surfaces of constant negative curvature.  Such 
surfaces are represented by a tessellation of the hyperbolic plane by regular polygons.  The Higgs field is given implicitly in terms of 
Schwarz triangle functions and analytic solutions are available for certain highly symmetric configurations.
\end{abstract}
\section{Introduction}
Consider the Abelian-Higgs field theory on a background surface $M$ with metric $ds^2=\Omega(x,y)\,(dx^2+dy^2)$.  At critical 
coupling the static energy functional satisfies a Bogomolny bound
\begin{equation}
E\,=\,\frac{1}{2}\int\left(\frac{B^2}{2\Omega}+|D_i\Phi|^2+\frac{\Omega}{2}\left(1-|\Phi|^2\right)^2\right)dxdy\,\geq\,\pi N,\label{YMHenergy}
\end{equation}
where the topological invariant $N$ (the `vortex number') is the number of zeros of $\Phi$ counted with multiplicity \cite{JT80}.  In the 
notation of \cite{Bap14} we have taken $e^2=\tau=1$.  Equality in \eqref{YMHenergy} is attained when the fields satisfy the Bogomolny 
vortex equations, which are obtained by completing the square in \eqref{YMHenergy}.  In complex coordinates $z=x+\text{i}y$ these are
\begin{equation}
D_{\bar{z}}\Phi\,=\,0,\qquad\qquad B\,=\,\Omega\,(1-|\Phi|^2).\label{bogomolny}
\end{equation}
This set of equations has smooth vortex solutions.  As we explain in section \ref{introduction}, analytical results are most readily 
obtained when $M$ is hyperbolic, having constant negative curvature $K=-1$, a case which is of interest in its own right due to the relation between hyperbolic 
vortices and $\text{SO}(3)$-invariant instantons, \cite{Wit77}.

The aim of this paper is to construct solutions to the vortex equations on {\it compact} surfaces $M$ whose universal cover 
has a hyperbolic metric.  In section \ref{hypertilings} we will see how such surfaces can be obtained as quotients of 
the hyperbolic plane.  The problem of finding one vortex on $M$ is then equivalent to finding a regular lattice of vortices in the hyperbolic plane, 
which may be of interest as a hyperbolic version of the Abrikosov vortex lattice observed experimentally in 
superconductors \cite{ET67}.

Our approach involves the construction of the Higgs field $\Phi$ from a holomorphic function $f(z)$ satisfying certain periodicity 
conditions.  The bulk of this paper (section \ref{solutions}) is concerned with constructing this map for an especially 
symmetric genus $2$ surface, the Bolza surface.  Section \ref{highergenus} extends the construction to certain higher genus surfaces and 
we wrap up in section \ref{conclusions} with some ideas for future work.

\section{Hyperbolic vortices}\label{introduction}
The Bogomolny equations \eqref{bogomolny} for Abelian-Higgs vortices at critical coupling on a curved surface can be combined to 
give the Taubes equation
\begin{equation}
\nabla^2\log|\Phi|^2+2\,\Omega(1-|\Phi|^2)\,=\,4\pi\sum_{i=1}^N\delta^{(2)}(z-z_i).\label{vortexequation}
\end{equation}
The right hand side provides sources for the vortices, where the Higgs field $\Phi$ vanishes.  The Higgs field can be written as $\Phi=\text{e}^{h/2+\text{i}\chi}$ with $h$ and $\chi$ real functions 
of $z$ and $\bar{z}$.  The phase $\chi$ depends on the gauge choice and increases by $2\pi$ around a unit charge vortex.  
Expressed in terms of these functions, the components of the $1$-form gauge potential and magnetic flux density are
\begin{equation}
a_{\bar{z}}\,=\,a_z^\ast\,=\,-\text{i}\partial_{\bar{z}}\left(\tfrac{1}{2}h+\text{i}\chi\right),\quad\qquad B\,=\,-2\text{i}F_{z\bar{z}}\,=\,-2\text{i}\left(\partial_za_{\bar{z}}-\partial_{\bar{z}}a_z\right)\,=\,-\tfrac{1}{2}\nabla^2h,\label{gaugepotential}
\end{equation}
where $\nabla^2=4\partial_z\partial_{\bar{z}}$ is the naive Euclidean Laplacian.
\\ \\
It was first observed by Witten, \cite{Wit77}, that on a background of constant negative Gaussian curvature
\begin{equation}%
K(z,\bar{z})\,=\,-\frac{1}{2\Omega}\,\nabla^2\log(\Omega)\,=\,-1,
\end{equation}%
the Taubes equation becomes the Liouville equation and can be solved exactly.  We work with the Poincar\'e 
disk model of the hyperbolic plane, with metric
\begin{equation}%
ds^2\,=\,\frac{4}{(1-|z|^2)^2}\,dzd\bar{z}.
\end{equation}%
Geodesics for this metric are circular arcs which intersect the boundary of the unit disk at right angles, and 
the geodesic distance between two points is
\begin{equation}
d(z_2,z_1)\,=\,\cosh^{-1}\left(1+\frac{2\,|z_2-z_1|^2}{(1-|z_1|^2)(1-|z_2|^2)}\right).\label{hypdist}
\end{equation}
The general solution to the Taubes equation \eqref{vortexequation} constructed by Witten is a Higgs field
\begin{equation}
\Phi(z,\bar{z})\,=\,\frac{1-|z|^2}{1-|f(z)|^2}\,\frac{df}{dz},\label{generalvortex}
\end{equation}
so $|\Phi|^2$ is a ratio of two hyperbolic metrics of equal curvature \cite{Bap14},
\begin{equation}
|\Phi|^2\,=\,\frac{\tilde\Omega(f(z))}{\Omega(z)}\,\left|\frac{df}{dz}\right|^2.
\end{equation}
The function $w=f(z)$ is holomorphic with $|f|<1$ on the interior of the disk and $|f|=1$ on the boundary 
$|z|=1$, so $f:z\to w$ is a map from the Poincar\'e disk to itself.  The map has ramification points at the 
vortex positions, where, for a vortex of unit multiplicity, $\Phi$ has a simple zero and $f(z)$ behaves locally quadratically.  Solving the Taubes 
equation on hyperbolic space thus boils down to finding a holomorphic map $f:\mathbb{H}^2\to\mathbb{H}^2$ 
with ramification points of order one greater than the multiplicity of $\Phi$ at each vortex position.  The Schwarz-Pick lemma guarantees 
that $|\Phi|<1$ on the interior of the disk, with $|\Phi|=1$ everywhere on the boundary (where the Higgs field takes its vacuum value).

Let us consider the case where the domain of $f$ is a compact surface $(M,\Omega)$ of area $A_M$ and genus $g$.  Then the target surface $(\tilde M,\tilde\Omega)$ has a 
finite area which is obtained by pulling back the area form on $\tilde M$ by $f$.  Integrating the Taubes equation \eqref{vortexequation} 
over $M$ and using the topological invariant $\int\!B\,dxdy=2\pi N$ gives
\begin{equation}
A_{\tilde M}\,=\,\int_{\tilde M}dA_{\tilde{M}}\,=\,\int_M|\Phi|^2\,\Omega\,dxdy\,=\,\int_M\Omega\,dxdy-2\pi N\,=\,A_M-2\pi N,\label{amtilde}
\end{equation}
where $N$ is the number of zeros of $\Phi$ counted with multiplicity.  Positivity of $A_{\tilde M}$ implies the Bradlow bound on the 
number of vortices on a compact surface:
\begin{equation}
2\pi N\,<\,A_M\,=\,4\pi(g-1),\label{bradlow}
\end{equation}
where the relation between the area and genus of the surface $M$ follows from the Gauss-Bonnet theorem $\int\!(-K)dA=4\pi(g-1)$.  The vacuum solution with $N=0$ 
has $\Phi=1$ everywhere, while at the Bradlow limit $\Phi=0$.  The interpretation is that each vortex takes up an effective area 
$2\pi$.\footnote{The Bradlow bound \eqref{bradlow} can be modified to allow for an arbitary number of vortices on the surface.  By 
changing the potential term in \eqref{YMHenergy} to $\Omega(\tau-|\Phi|^2)^2/2$, the Bradlow inequality becomes $2\pi N<\tau A_M$.  
However, this has the effect of changing the curvature of the target space to $K_{\tilde M}=-\tau$, and the Taubes equation is no longer 
integrable.}

It is important to check that there is no contribution from the cross term arising when completing the square in \eqref{YMHenergy}.  This 
term is a total derivative and vanishes identically on a compact surface.

Multivaluedness of $f$ means that the area of each sheet of $\tilde M$ is in fact the value of $A_{\tilde M}$ given in \eqref{amtilde} 
divided by one plus the number of vortices on the interior of $M$.  This is in essence the Riemann-Hurwitz theorem and will be important 
in sections \ref{solutions} and \ref{highergenus}, where $\tilde M$ is considered as a tessellation of $\mathbb{H}^2$.

\subsection*{Example I: a finite number of vortices}
For a finite number of vortices, $N$, on $\mathbb{H}^2$, the function $f(z)$ is a Blaschke product \cite{Wit77}
\begin{equation}
f(z)\,=\,\prod_{j=1}^{N+1}\,\frac{z-a_j}{1-\bar{a}_jz},\qquad|a_j|\,<\,1\quad\forall j.\label{blaschkeproduct}
\end{equation}
In general the relation between the complex numbers $a_j$ and the vortex positions (at the critical points of $f$) is not obvious.  
Constructing a vortex solution from the vortex positions can only be done analytically for symmetric configurations, such as a cyclic 
ring of vortices.  The parameters needed for an arbitrary prescribed vortex arrangement must be found numerically.
\subsection*{Example II: a singly periodic array}
A vortex solution on the hyperbolic cylinder $\mathbb{H}^2/\mathbb{Z}$ was presented in \cite{MR10}.  The 
solution was constructed via a map $f$ from a half annulus in the upper half plane to a Euclidean rectangle, 
giving rise to a doubly periodic function.  In Poincar\'e disk coordinates, $f$ is given by a ratio 
of Jacobi elliptic functions,
\begin{equation}
f(z)\,=\,\frac{\text{cd}\Big(\frac{4\text{\bf{K}}}{\pi}\tan^{-1}(z);k\Big)-1}{\text{cd}\Big(\frac{4\text{\bf{K}}}{\pi}\tan^{-1}(z);k\Big)+1}.\label{Jacobi}
\end{equation}
Vortices (i.e.~zeros of $f'(z)$) are located at $z=\text{i}\tanh(n\lambda/4)$ with $\lambda=\pi\text{\bf{K}}'/\text{\bf{K}}$ and $n\in\mathbb{Z}$, where $\text{\bf{K}}(k)$ is the complete 
elliptic integral of the first kind with elliptic modulus $k$, and $\text{\bf{K}}'(k)=\text{\bf{K}}(\sqrt{1-k^2})$.  Notice that there is a double zero of 
$f(z)$ at every other zero of $\Phi$.  This suggests that $f$ can be expressed as an infinite Blaschke 
product
\begin{equation}
f(z)\,=\,z^2\,\prod_{j=1}^\infty\left(\frac{z^2-a_j^2}{1-a_j^2z^2}\right)^2\qquad\text{with}\qquad a_j\,=\,\text{i}\tanh\left(\frac{j\lambda}{2}\right).\label{infprod}
\end{equation}
Equality between \eqref{Jacobi} and \eqref{infprod} is easily checked at $z=\text{i}\tanh(n\lambda/4)$ by known infinite products.  The 
{\it Blaschke condition}
\begin{equation}
\sum_{j=1}^\infty(1-|a_j|)\,<\,\infty\label{blaschkecondition}
\end{equation}
holds for $\lambda>0$ (the sum can be evaluated as a $q$-hypergeometric function), thereby ensuring convergence of the product \eqref{infprod}.

Using the closed form expression \eqref{Jacobi}, the Higgs field \eqref{generalvortex} near the vortex at the origin and 
at the midpoint between vortices is
\begin{equation}%
|\Phi|_{\text{origin}}\,=\,\frac{8\text{\bf{K}}^2}{\pi^2}\,(1-k^2)\,|z|+\dots,\qquad\qquad|\Phi|_{\text{midpoint}}\,=\,\frac{2\text{\bf{K}}}{\pi}\,(1-k).
\end{equation}%
Periodicity of $|\Phi|$ follows from the fact that a period shift in $z$ is equivalent to a fractional linear 
transformation of $f$:
\begin{equation}%
z\,\mapsto\,z'\,=\,\frac{z-\text{i}\tanh(\lambda/4)}{\text{i}z\tanh(\lambda/4)+1}\qquad\Rightarrow\qquad f(z')\,=\,\frac{(1+k)f(z)-(1-k)}{(1-k)f(z)-(1+k)},
\end{equation}%
a symmetry which will be discussed in more detail in section \ref{fproperties}.

\section{Hyperbolic tessellations}\label{hypertilings}
The hyperbolic plane has infinitely many tessellations by regular polygons whose edges are 
geodesic arcs.  Such tessellations are denoted by their Schl\"afli symbol $\{p,q\}$, where $p$ is the number of sides 
of the fundamental polygon and $q$ the number of such polygons meeting at each vertex.  These polygons tile the hyperbolic plane if $q(p-2)>2p$ and, in 
contrast to the case of tessellations of the Euclidean plane, have a fixed area
\begin{equation}
A_{\{p,q\}}=\frac{\pi}{(-K)}\left(p-2-\frac{2p}{q}\right),\label{areaformula}
\end{equation}
where in our conventions $K=-1$.

We follow \cite{FP87} in defining the Euler characteristic of any regular tessellation to be
\begin{equation} 
\chi_{\{p,q\}}=-A_{\{p,q\}}/2\pi.\label{chipq}
\end{equation}
As discussed below, a given tessellation describes a smooth compact surface of genus $g$ if $\chi$ is a negative even integer, 
$\chi=2-2g$.  The surface is obtained by identifying pairs of edges of the fundamental polygon.
\\ \\
The uniformization theorem guarantees that a compact surface $M$ of genus $g\geq2$ possesses 
a global hyperbolic metric.  The universal covering space of such a surface is the hyperbolic plane, 
with a tessellation whose fundamental domain is obtained as the quotient $\mathbb{H}^2/H$, where 
$H\cong\pi_1(M)$ is the first homotopy group of the surface.  The usual 
application of the Gauss-Bonnet theorem tells us that the area of a compact genus $g$ surface with curvature $K=-1$ is 
$A=-2\pi\chi=4\pi(g-1)$.  Note from \eqref{areaformula} that a tessellation with $\{p,q\}=\{4g,4g\}$ has the correct area for a surface 
of genus $g$.

For given $g$, there is a $3(g-1)$-complex-dimensional family of surfaces parametrised by the 
lengths of $3(g-1)$ non-intersecting closed geodesics on the fundamental polygon, together with `twist' data 
describing the gluing of the edges.  These parameters are known as the Fenchel-Nielsen coordinates 
on Teichm\"uller space, \cite{Naz}.  For ease of computation we specialise to the most symmetric 
surfaces, corresponding to tessellations by regular polygons.  As an example, a smooth genus $2$ surface 
can be constructed from any of the tessellations
\begin{equation*}
\{8,8\},\quad\{10,5\},\quad\{12,4\},\quad\{18,3\}.
\end{equation*}
Of the tessellations above, our focus will be mainly on the highly symmetric tessellation by regular octagons in 
the $\{8,8\}$ configuration with opposite edges identified.  Such a tessellation is obtained as the 
quotient of the unit disk by the freely acting subgroup $H\subset\Gamma$ of the full automorphism 
group $\Gamma$ of the tessellation.  $H$ is a Fuchsian group generated by eight hyperbolic M\"obius transformations 
$M_k$.  This is a discrete subgroup of $\mathrm{PSL}(2,\mathbb{C})$ 
with the fixed-point-free condition $(\text{tr}M_k)^2>4$.  Note that $\Gamma$ also contains elliptic 
elements with fixed points in the interior of the disk; these elements are excluded from the quotient 
in order to obtain a smooth surface.  If we orient the tessellation such that there are octagon centres at the origin and 
along the coordinate axes, then the eight M\"obius transformations $M_k(z)$ with $k=0,\dots,7$ take the form
\begin{equation}
M_k(z)\,=\,\frac{z+L\text{e}^{\text{i}k\pi/4}}{L\text{e}^{-\text{i}k\pi/4}z+1}\qquad\text{with}\qquad L\,=\,\sqrt{2\sqrt{2}-2}\,\approx\,0.910,\label{MT}
\end{equation}
where $L$ is the Euclidean distance from the origin to one of the neighbouring octagon centres and we have the relation 
$M_{k+4}=M_k^{-1}$.  This identification of the tessellation amounts to defining the fundamental octagon, 
$\mathcal{O}$, by
\begin{equation}
\mathcal{O}\,=\,\{x\in\mathbb{H}^2\,|\,d(x,0)\leq d(x,M_k(0)),\,k=0,\dots,7\},\label{fundcell}
\end{equation}
where the hyperbolic distance function was defined in \eqref{hypdist}.  Note that hyperbolic tessellations 
do not, in general, correspond to compact surfaces, but a fundamental polygon 
can still be defined as in \eqref{fundcell}, and the tessellation has a fractional Euler characteristic 
defined by \eqref{chipq}.

\begin{figure}
\begin{minipage}{0.485\linewidth}
\centering
\includegraphics[width=0.73\linewidth]{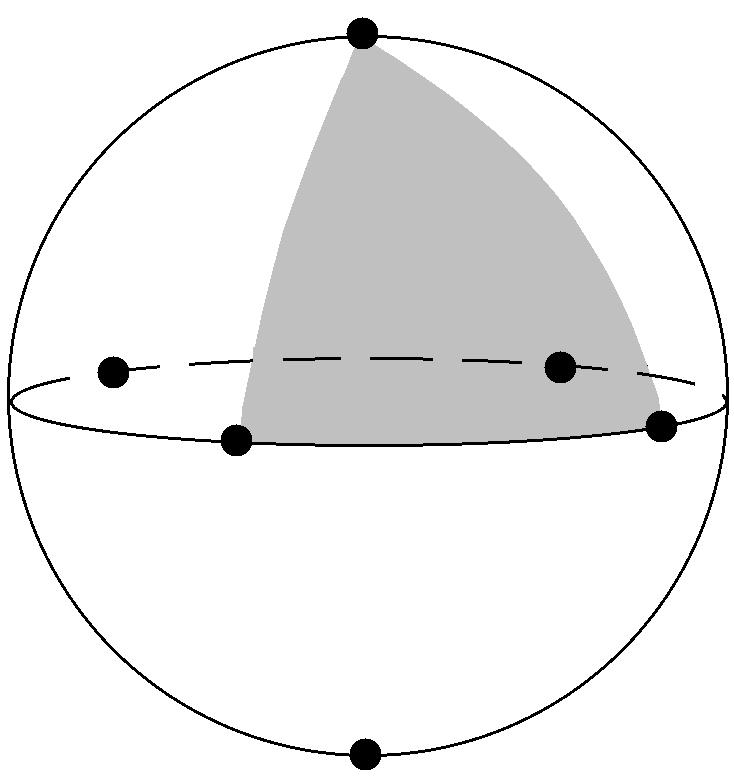}
\end{minipage}
\begin{minipage}{0.485\linewidth}
\centering
\includegraphics[width=0.75\linewidth]{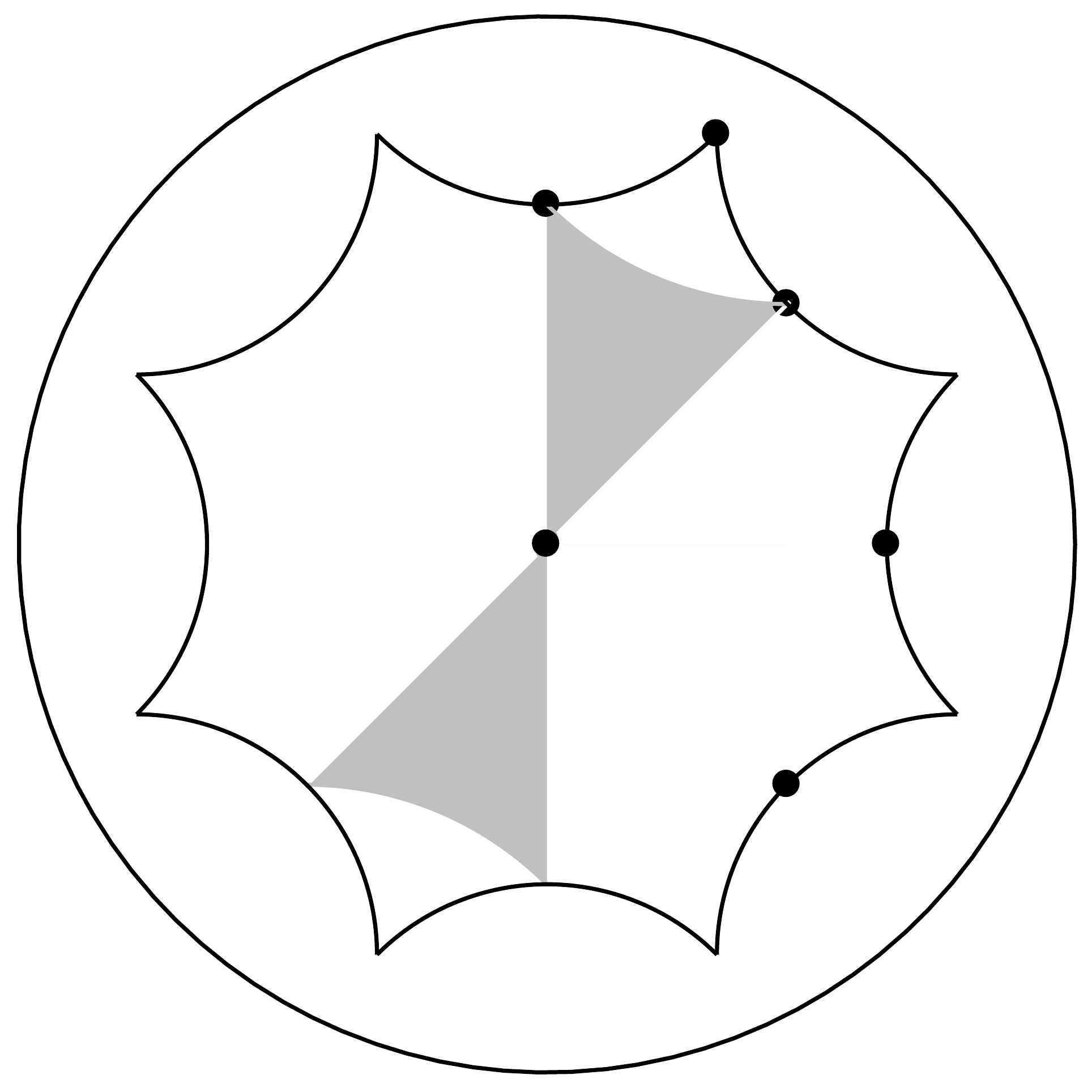}
\end{minipage}
\caption{The Bolza surface can be represented as a double cover of the Riemann sphere branched at the vertices of an inscribed 
regular octahedron (left), or equivalently as an $\{8,8\}$ tessellation of the hyperbolic plane (right) where opposite 
edges of the fundamental octagon are identified.  There is a $C_8$ symmetry at each of the branch points, which are 
indicated by dots.  The shaded equilateral triangle with angles $\pi/2$ on the sphere is covered by the two shaded equilateral triangles 
with angles $\pi/4$ on the Bolza surface (one for each branch).  The north and south poles of the sphere correspond to the centre and 
vertex of the octagon, and hence to antipodal points on the Bolza surface.}\label{figoct}
\end{figure}

The fundamental polygon of the regular $\{8,8\}$ tessellation with opposite edges identified is the most symmetric genus $2$ surface, 
known as the Bolza surface.  It has an automorphism group of order $48$ ($96$ if one includes reflections) and can be thought of as a 
double cover of $\mathbb{CP}^1$ with ramification points at the six vertices of an inscribed regular octahedron.  On the hyperbolic 
octagon, the ramification points correspond to points of $C_8$ symmetry.  There are also $16$ points with $C_3$ symmetry and $24$ with 
$C_2$ symmetry, as can be seen by dividing the octagon into $96$ triangles with angles $(\tfrac{\pi}{2},\tfrac{\pi}{3},\tfrac{\pi}{8})$.  
Most of this symmetry is broken when a vortex is placed on the surface.  However, when the vortex is placed at the origin, the $C_8$ 
symmetry about both the origin and the vertex is preserved, which will be key to our analytical construction.  The fundamental domain for 
the $\{8,8\}$ tessellation, together with the positions of the symmetric points, is sketched in figure \ref{figoct}.  As a hyperelliptic 
curve, the genus $2$ Bolza surface can be expressed as
\begin{equation}%
y^2\,=\,x(x^4+1).
\end{equation}%
We will return to the representation of the Bolza surface as a hyperelliptic curve in the context of compact surfaces of higher genus in 
section \ref{highergenus}.

\newpage
\section{A vortex on the Bolza surface}\label{solutions}
\subsection{Identifying the holomorphic map}\label{fproperties}
The aim of this section is to compute the Higgs field of a vortex placed at the centre of the period octagon.  As described in section \ref{introduction}, we 
proceed by looking for a holomorphic function $f:\mathbb{H}^2\to\mathbb{H}^2$ ramified at the centre of every octagonal cell in the 
tessellation.  Restricted to the fundamental octagon, the map $w=f(z)$ is conformal except at the vortex position, where angles are 
doubled.  The image is precisely a square 
in the $\{4,8\}$ tessellation, and the map $f$ is two-to-one, providing a double 
cover of the square with double winding around the centre.  Pairs of identified opposite edges of the octagon map to the 
same edge of the square.  It should be noted that opposite edges of the square are not identified, as the $\{4,8\}$ tessellation 
does not correspond to a compact surface.  A M\"obius transformation \eqref{MT} mapping one vortex to the next corresponds to a rotation 
about one of the four edge centres of the fundamental square,
\begin{equation}
\tilde{M}_k(f(z))\,=\,f(M_k(z))\,=\,\frac{\sqrt{2}f(z)-L\text{e}^{\text{i}k\pi/2}}{L\text{e}^{-\text{i}k\pi/2}f(z)-\sqrt{2}},\label{fMT}
\end{equation}
which we deduce by considering the effect of $f$ on the vertices of the octagon.

The fact that $f$ transforms by a M\"obius transformation shows that it is {\it not}, as might have been expected, an automorphic form.  
Instead, if $z$ and $f(z)$ both transform by fractional linear transformations
\begin{equation}%
z\,\mapsto\,\frac{bz+c}{\bar{c}z+\bar{b}},\qquad\qquad f(z)\,\mapsto\,\frac{\beta f(z)+\gamma}{\bar{\gamma}f(z)+\bar{\beta}},
\end{equation}%
with $|b|^2-|c|^2=1=|\beta|^2-|\gamma|^2$, then $\frac{df}{dz}$ transforms as
\begin{equation}
\frac{df}{dz}\,\mapsto\,\left(\frac{\bar{c}z+\bar{b}}{\bar{\gamma}f(z)+\bar{\beta}}\right)^2\frac{df}{dz}.\label{dfdz}
\end{equation}
For the specific case in question, with the fractional linear transformations \eqref{MT} and \eqref{fMT}, we have
\begin{equation}%
\frac{df}{dz}\,\mapsto\,(2\sqrt{2}+4)\left(\frac{L\text{e}^{-\text{i}k\pi/4}z+1}{L\text{e}^{-\text{i}k\pi/2}f(z)-\sqrt{2}}\right)^2\,\frac{df}{dz}.
\end{equation}%
The factor multiplying $\frac{df}{dz}$ can be thought of as a gauge transformation of the vortex fields, which does not affect $|\Phi|$.  Despite this complicated transformation 
property of $f$ and $\frac{df}{dz}$, we will see in section \ref{mappingbetweenpolygons} that automorphic forms do in fact appear in our computation of the vortex fields.

The area of the image square follows from the Riemann-Hurwitz theorem.  Denote by $A$ and $A'$ the 
areas of the fundamental polygons in the domain and image tessellations, which are related to the Euler 
characteristic of the tessellation via \eqref{chipq}.  The map $f$ has a single ramification point of order $2$, so 
$A'=\tfrac{1}{2}(A-2\pi)$ in agreement with \eqref{areaformula}, which gives $A_{\{8,8\}}=4\pi$ and $A'_{\{4,8\}}=\pi$.  More generally, 
if $N$ vortices are placed in the interior of a domain polygon representing a compact genus $g$ surface, then the map $f$ has $N$ 
ramification points and is of degree $N+1$.  By the Riemann-Hurwitz theorem, the area of the image polygon is given by
\begin{equation}
A'\,=\,\frac{1}{N+1}\,(A-2\pi N)\,=\,\frac{2\pi}{N+1}\,(2g-2-N).\label{bbradlow}
\end{equation}
The requirement that $A'>0$ implies that the number of vortices permitted on a genus $g$ surface is constrained by
\begin{equation}
N\,<\,2g-2,\label{bradlowagain}
\end{equation}
the Bradlow bound \eqref{bradlow} again.  The map we have described between the octagon and the square can hence be generalised to more 
vortices on a higher genus surface (see also sections \ref{comphiggsfield} and \ref{highergenus}).
\subsubsection*{Non-existence of Blaschke product}
Before proceeding any further, we should investigate whether the map $f$ can be expressed as an infinite Blaschke product with 
zeros of $f$ occurring at the same density as vortices, mimicking the way this was done for the 
one-dimensional vortex array (\ref{Jacobi}, \ref{infprod}).  For $f$ of the form \eqref{blaschkeproduct} with $N\to\infty$, 
this will be the case if the Blaschke condition \eqref{blaschkecondition} holds.

Consider a circle $|z|=r$ in the Poincar\'e disk with $0\ll r<1$.  As each octagon in the tessellation has area $4\pi$, the interior of 
the circle 
will contain $N(r)=A(r)/4\pi=r^2/(1-r^2)$ vortices.  Then the number of vortices in an annulus of thickness $dr$ is $dN=2rdr/(1-r^2)^2$.  
Using this measure, the sum \eqref{blaschkecondition} can be converted to an integral if we assume that the $\{|a_j|\}$ 
have the density of the tessellation:
\begin{equation}%
\sum_{j=1}^\infty(1-|a_j|)\,\sim\,\lim_{\epsilon\to0}\int^{1-\epsilon}(1-r)\,\frac{2r}{(1-r^2)^2}\,dr.
\end{equation}%
The right hand side diverges logarithmically, hence there is no convergent Blaschke product with the symmetries of the octagonal 
tessellation.  For the hyperbolic cylinder discussed in section \ref{introduction} the number of vortices as a function of 
distance from the origin is $N(r)\propto\tanh^{-1}r$, so $dN\propto dr/(1-r^2)$ and the corresponding integral converges.

\subsection{Mapping between polygons}\label{mappingbetweenpolygons}
\begin{figure}
\centering
\includegraphics[width=0.9\linewidth]{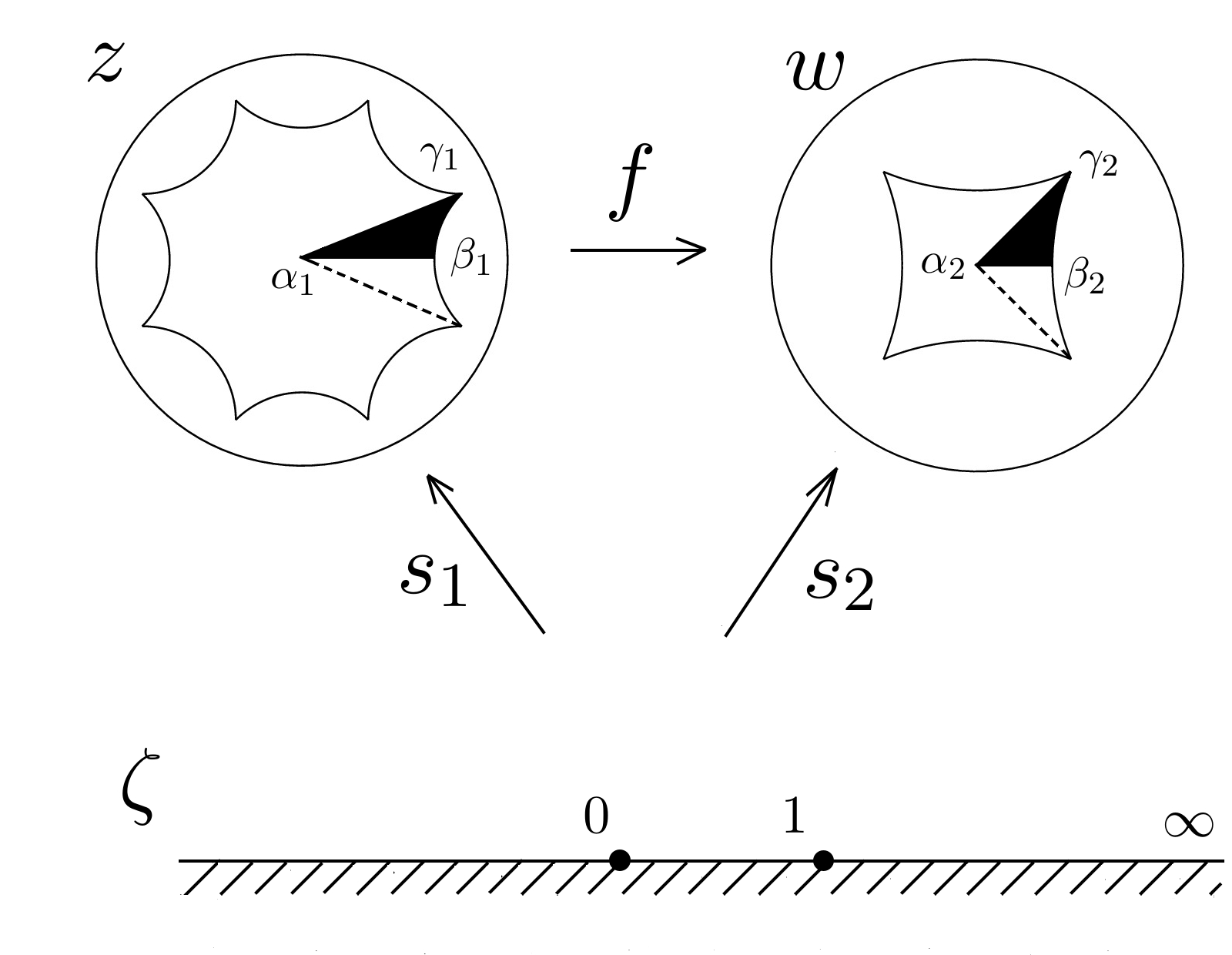}
\caption{Sketch showing the action of the map $f$ on a triangle $\mathcal{O}_\Delta$ within the fundamental octagon $\mathcal{O}$, via an 
auxiliary upper half plane.  The image square and triangle are denoted $\mathcal{S}$ and $\mathcal{S}_\Delta$.  In our conventions, 
$\zeta=0$ maps to the vertex of the shaded triangles at the origin of the tessellations, while $\zeta=1$ and $\zeta=\infty$ map to the 
other two vertices, respectively the edge midpoint and vertex of the polygons.  The real axis of the upper half plane maps to the 
boundary of the triangle.  The angles of the black triangles are $(\pi\alpha_1,\pi\beta_1,\pi\gamma_1)=(\tfrac{\pi}{8},\tfrac{\pi}{2},\tfrac{\pi}{8})$ 
and $(\pi\alpha_2,\pi\beta_2,\pi\gamma_2)=(\tfrac{\pi}{4},\tfrac{\pi}{2},\tfrac{\pi}{8})$.  The Higgs field at a vertex of the octagon is 
easier to compute if we instead use the dashed triangles, which has the effect of swapping $\beta$ and $\gamma$.}\label{mapf}
\end{figure}

The map $f$ between the polygonal tessellations of the Poincar\'e disk described in section \ref{fproperties} is easy to implement because 
the position of the vortex is consistent with a regular tiling of the fundamental polygon by congruent triangles with a shared vertex at the vortex 
position.  This procedure avoids mapping directly between the octagon and the square, which would introduce a large number of additional 
{\it accessory parameters} which would then need to be determined.  We thus divide the fundamental octagon into $16$ triangles, each of which maps, via 
an auxiliary half plane, to an eighth of the image square.  The fundamental domains $\mathcal{O}_\Delta$ (with Poincar\'e disk coordinate 
$z$) and $\mathcal{S}_\Delta$ (with coordinate $w$) are shown in figure \ref{mapf}.  By Schwarz reflection, the hyperbolic plane is 
covered by repeated reflection of the triangle $\mathcal{O}_\Delta$ in its edges.

The existence of the maps in figure \ref{mapf} is guaranteed by the Riemann mapping theorem.  In the following paragraphs we sketch how 
the Riemann map is constructed for curvilinear polygons before specialising to the case of triangles, where one can obtain explicit 
results.  Then in section \ref{comphiggsfield} we will focus on the specific triangles in figure \ref{mapf}.
\subsubsection*{The Riemann map}
The Riemann mapping theorem states the existence of a unique conformal map between the upper half plane and the interior of a simply 
connected domain of $\mathbb{C}$.  Specialising to the case of an $n$-sided polygon bounded by circular arcs, the map is bijective 
and extends to a map from the boundary of the upper half plane to the boundary of the polygon.  It is conformal everywhere except at the 
$n$ points $\zeta_i$ on the real axis which map to the vertices of the polygon.  The requirement that the image of the map 
should extend to the entire Poincar\'e disk by a series of reflections leads to an $\text{SL}(2,\mathbb{R})$-invariant differential 
equation satisfied by the function $s$ representing the map.  This third order non-linear ordinary differential equation is solved by a 
ratio of linearly independent solutions to a second order linear ordinary differential equation with $n$ regular singular points at $\zeta=\zeta_i$, 
where the exponents depend on the angle at each corresponding vertex of the polygon.  The 
interested reader is referred to \cite{Neh52} for details.  From now on we specialise to the case of triangles, in which case the Riemann map is 
known as the Schwarz triangle map.
\subsubsection*{The Schwarz triangle map}
In order to make use of the Riemann mapping theorem we introduce an auxiliary upper half plane with complex coordinate $\zeta$.  Then 
there are two Schwarz triangle maps, $s_1:\zeta\to z$ and $s_2:\zeta\to w$, mapping the upper half plane to each of the triangles in 
figure \ref{mapf}.  The map $f$ between the triangles $\mathcal{O}_\Delta$ and $\mathcal{S}_\Delta$ is given by the composition 
$w=f(z)=s_2(s_1^{-1}(z))$.  Note that when the images are extended to the entirety of the Poincar\'e disk (by reflections both in the 
triangle edges and from the auxiliary upper half plane to the lower half plane), the functions $s_1(\zeta)$ and 
$s_2(\zeta)$ are multivalued.  On the other hand, the inverse maps $s_1^{-1}(z)$ and $s_2^{-1}(w)$ are automorphic forms invariant under 
the elements of the automorphism group of the octagon, and are dense in both zeros and poles as the boundary of the 
disk is approached.  In general there is no closed form expression for $s_1^{-1}(z)$, so we use a parametric approach:~picking an arbitrary 
point on the $\zeta$ plane will tell us the value of $f(z)$ at $z=s_1(\zeta)$.  As we shall see, there are some particular points on 
$\mathcal{O}_\Delta$ at which we can expand $f(z)$ as a series, allowing $\Phi(z)$ to be computed analytically using \eqref{generalvortex}.

By convention, we take the points $\zeta_i=0,1,\infty$ on the upper half plane to map to the vertices of a triangle with angles 
$\pi\alpha$, $\pi\beta$, $\pi\gamma$, respectively.  The map from the upper half plane to this triangle is given by a ratio of two 
linearly independent solutions of the hypergeometric equation \cite{Neh52}
\begin{equation}
\frac{d^2y}{d\zeta^2}+\left(\frac{c}{\zeta}+\frac{d}{\zeta-1}\right)\frac{dy}{d\zeta}+\frac{ab}{\zeta(\zeta-1)}\,y\,=\,0.\label{riemannde}
\end{equation}
The two solutions at $\zeta=0$ have exponents $0$ and $(1-c)$, and are given by the hypergeometric functions
\begin{equation}%
y(\zeta)\,=\,F(a,b;c;\zeta)\qquad\qquad\tilde{y}(\zeta)\,=\,\zeta^{1-c}F(a',b';c';\zeta),
\end{equation}%
where $a$, $b$, $c$ and $d$ are related to the angles of the triangle by
\begin{equation}
\left\{\quad\begin{array}{rclcrcl}a&=&\tfrac{1}{2}(1-\alpha-\beta+\gamma)&\qquad&a'&=&a-c+1\\
b&=&\tfrac{1}{2}(1-\alpha-\beta-\gamma)&\qquad&b'&=&b-c+1\\
c&=&1-\alpha&\qquad&c'&=&2-c\\
d&=&1-\beta\,=\,a+b-c+1&\qquad&d'&=&d.
\end{array}\right.\label{abcd}
\end{equation}
The Schwarz triangle function is then defined as
\begin{equation}%
s(\zeta)\,=\,\mathcal{N}\,\frac{\tilde{y}(\zeta)}{y(\zeta)},
\end{equation}%
where the normalisation factor $\mathcal{N}$ is chosen so as to ensure that the triangle has the correct size, such that geodesic 
extensions of the circular edges intersect the unit disk at right angles.  This calculation was performed in \cite{HM03}.  Putting 
everything together and simplifying, we have
\begin{equation}
s(\zeta)\,=\,\sqrt{\frac{\sin(\pi a')\sin(\pi b')}{\sin(\pi a)\sin(\pi b)}}\,\frac{\Gamma(a')\Gamma(b')\Gamma(c)}{\Gamma(a)\Gamma(b)\Gamma(c')}\,\zeta^{1-c}\,\frac{F(a',b';c';\zeta)}{F(a,b;c;\zeta)}.\label{trianglemap}
\end{equation}

\subsection{Computing the Higgs field}\label{comphiggsfield}
In order to calculate the Higgs field, we use the two triangle maps $s_1(\zeta)$ and $s_2(\zeta)$ as 
depicted in figure \ref{mapf}.  The derivative of $f$ is obtained by the chain rule.  Then substituting 
into \eqref{generalvortex} gives
\begin{equation}
\Phi(z)\,=\,\frac{1-|s_1(\zeta)|^2}{1-|s_2(\zeta)|^2}\,\frac{ds_2(\zeta)}{d\zeta}\left(\frac{ds_1(\zeta)}{d\zeta}\right)^{-1}.\label{Phisol}
\end{equation}
Recall that $(\alpha,\beta,\gamma)$ are the angles in the triangle (divided by $\pi$), and $(a,b,c)$ are defined through \eqref{abcd}.  
The triangles of interest (figure \ref{mapf}) have $(\alpha_1,\beta_1,\gamma_1)=(\frac{1}{8},\frac{1}{2},\frac{1}{8})$ and 
$(\alpha_2,\beta_2,\gamma_2)=(\frac{1}{4},\frac{1}{2},\frac{1}{8})$.  The Higgs field of a vortex placed at the origin can be computed analytically at the 
vertices of the fundamental triangle $\mathcal{O}_\Delta$.  In order to do this, we expand $s(\zeta)$ near the singular points:
\begin{IEEEeqnarray}{rcl}
s(\delta)\,&=&\,\sqrt{\frac{\sin(\pi a')\sin(\pi b')}{\sin(\pi a)\sin(\pi b)}}\,\frac{\Gamma(a')\Gamma(b')\Gamma(c)}{\Gamma(a)\Gamma(b)\Gamma(c')}\,\delta^{\alpha}+\dots\label{szero}\\
s(1+\delta)&=&\sqrt{\frac{\sin(\pi a)\sin(\pi b)}{\sin(\pi a')\sin(\pi b')}}\left(1-\delta^{\beta}\,\frac{\sin(\pi\alpha)}{\pi\beta\Gamma^2(\beta)}\,\Gamma(1-a)\Gamma(1-b)\Gamma(1-a')\Gamma(1-b')+\dots\right)\nonumber\\\label{sone}
\end{IEEEeqnarray}
The formula \eqref{Phisol} then gives the Higgs field at the points on the octagon corresponding to the vertices of 
$\mathcal{O}_\Delta$.  Near $\zeta=0$ we have from \eqref{szero}
\begin{equation}%
z\,=\,s_1(\zeta)\,=\,\mathcal{N}_1\zeta^{1/8}+\dots,\qquad\qquad s_2(\zeta)\,=\,\mathcal{N}_2\zeta^{1/4}+\dots.
\end{equation}%
Using \eqref{Phisol} and the expansion of $s_1(\zeta)$ to change from $\zeta$ to $z$ coordinates gives
\begin{equation}
\Phi\,=\,2\,\frac{\mathcal{N}_2}{\mathcal{N}_1}\,\zeta^{1/8}+\dots\,=\,2\,\frac{\mathcal{N}_2}{\mathcal{N}_1^2}\,z+\dots.\label{Phiatorigin}
\end{equation}
At the vertex of $\mathcal{O}_\Delta$ corresponding to $\zeta=1$, the expansion is of the form
\begin{equation}%
s_1(1+\delta)\,=\,\mathcal{A}_1+\mathcal{B}_1\delta^{1/2}+\dots,\qquad\qquad s_2(1+\delta)\,=\,\mathcal{A}_2+\mathcal{B}_2\delta^{1/2}+\dots,
\end{equation}%
with coefficients that can read off from \eqref{sone}.  Applying \eqref{Phisol} gives
\begin{equation}%
\Phi\,=\,\frac{1-\mathcal{A}_1^2}{1-\mathcal{A}_2^2}\,\frac{\mathcal{B}_2}{\mathcal{B}_1}.\label{Phiatvertex}
\end{equation}%
Evaluating \eqref{Phiatorigin} and \eqref{Phiatvertex} for the triangles of interest, we find
\begin{IEEEeqnarray}{rcll}
|\Phi|_{\text{origin}}\,&=&\,\left[(4\pi)^{-3/2}\,\sin\!\left(\tfrac{\pi}{8}\right)\Gamma^2\!\left(\tfrac{1}{8}\right)\Gam{1}{4}\right]|z|+\mathcal{O}(|z|^3)&\,\approx\,1.768|z|+\mathcal{O}(|z|^3),\label{PhiexactO}\\
&&&\nonumber\\
|\Phi|_{\text{mid edge}}\,&=&\,\frac{\sqrt{2}\,\,\Gam{1}{8}\Gamma^2\!\left(\tfrac{1}{4}\right)\Gam{3}{8}}{\Gam{1}{16}\Gam{3}{16}\Gam{5}{16}\Gam{7}{16}}&\,\approx\,0.752,\label{Phiexact}\\
&&&\nonumber\\
|\Phi|_{\text{vertex}}\,&=&\,2^{-1/4}&\,\approx\,0.841.
\end{IEEEeqnarray}An expansion of the hypergeometric functions about the point at infinity can be avoided by a redefinition of the fundamental triangle, as 
explained in the caption to figure \ref{mapf}, thereby allowing the Higgs field at the vertex to be computed using \eqref{sone} and \eqref{Phiatvertex}, albeit with different values of the parameters.

\begin{figure}
\centering
\includegraphics[width=0.9\linewidth]{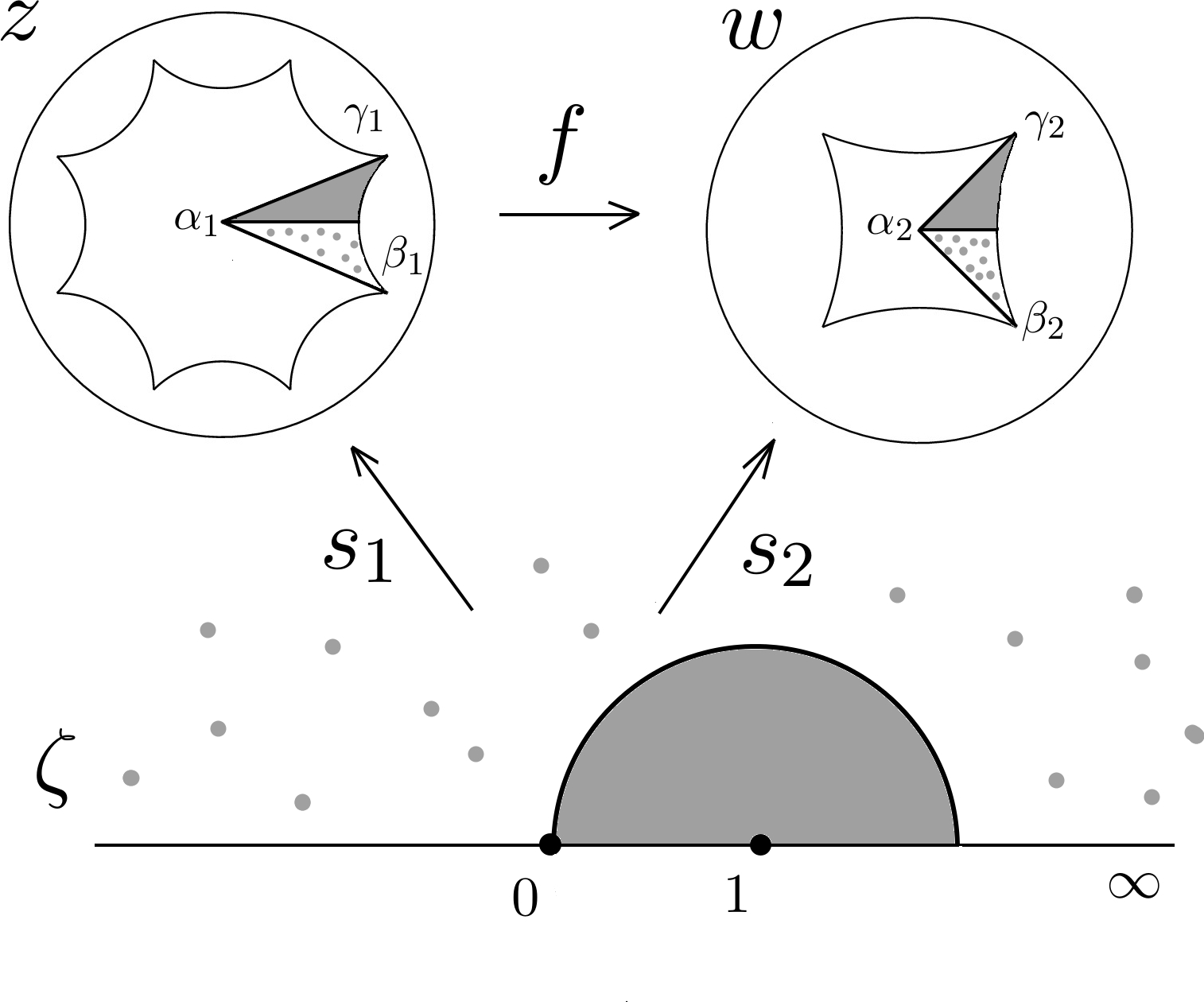}
\caption{An alternative definition of the map $f$ to that shown in figure \ref{mapf}.  The resulting 
Higgs field is unchanged by this map.  Shading indicates regions which map into each other, 
and the remainder of the hyperbolic plane is covered by reflections in the sides of the fundamental 
triangles.  The advantage of this redefinition is improved numerical sampling (as the parameter 
$\zeta$ need only take values in a semi-disk), although it is harder to compute the value of $|\Phi|$ 
at the midpoint of an edge analytically as in \eqref{Phiexact}.    The angles of the triangles are 
$(\pi\alpha_1,\pi\beta_1,\pi\gamma_1)=(\tfrac{\pi}{4},\tfrac{\pi}{8},\tfrac{\pi}{8})$ and 
$(\pi\alpha_2,\pi\beta_2,\pi\gamma_2)=(\tfrac{\pi}{2},\tfrac{\pi}{8},\tfrac{\pi}{8})$.}\label{mapf2}
\end{figure}

\begin{figure}
\begin{minipage}{0.485\linewidth}
\centering
\vspace{-0.3cm}
\includegraphics[width=0.75\linewidth]{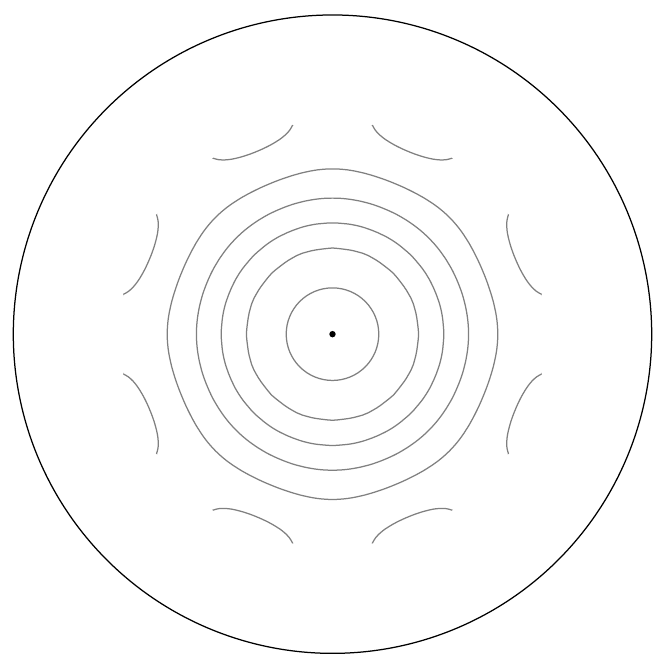}
\end{minipage}
\begin{minipage}{0.485\linewidth}
\centering
\includegraphics[width=0.75\linewidth]{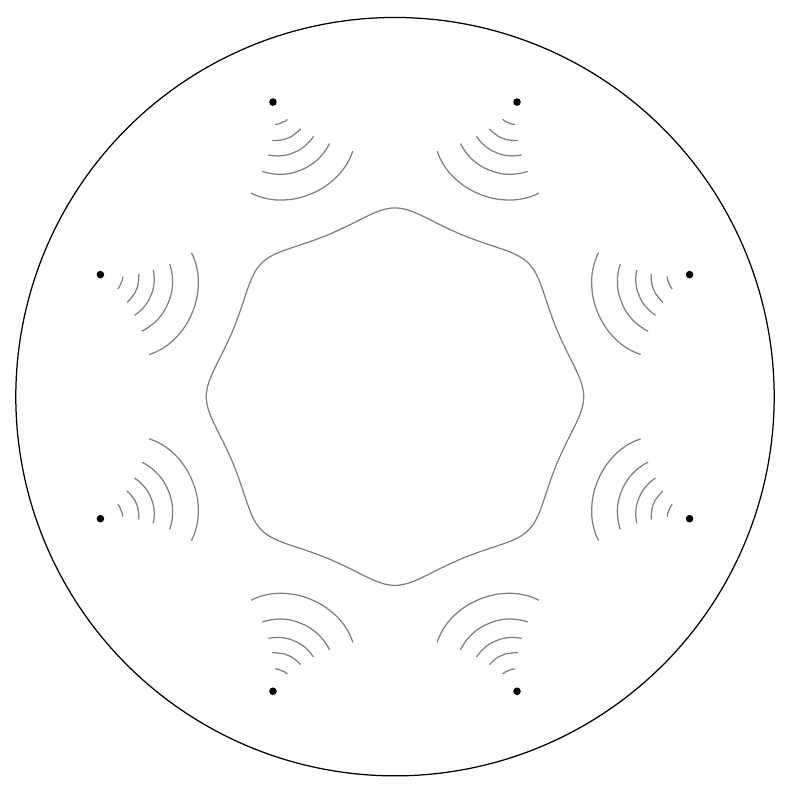}
\end{minipage}
\caption{Contour plots of $|\Phi|^2$ for a vortex at the centre of the fundamental octagon (left) and 
at a vertex (right).  $|\Phi|^2$ has a zero at the vortex position (marked by black dots) and reaches 
a maximum value of $1/\sqrt{2}$.  The two configurations are equivalent due to the two preserved $C_8$ 
symmetries of the fundamental octagon (figure \ref{figoct}).  The leading term of $|\Phi|^2$ near 
the centered vortex is at order $|z|^2$ and is circularly symmetric, while non-trivial angular dependence enters at order $|z|^{10}$.}\label{contours}
\end{figure}

Away from the points where the behaviour of the Higgs field can be studied analytically by known 
series expansions of the hypergeometric function, contours of $|\Phi|^2$ are obtained numerically and plotted using the 
auxiliary coordinate $\zeta$ as a parameter.  This method leads to difficulty when 
sampling in the vicinity of the vertex of the octagon, since this is the image of the point at 
infinity in the $\zeta$ plane.  A more effective procedure is to divide the octagon into eighths rather 
than sixteenths, so $(\alpha_1,\beta_1,\gamma_1)=(\tfrac{1}{4},\tfrac{1}{8},\tfrac{1}{8})$.  
The preimage of this triangle under $s_1(\zeta)$ is the entire upper half plane.  Splitting the triangle into 
two further isosceles triangles as shown in figure \ref{mapf2}, we find we need only sample within the 
region $|\zeta-1|\leq 1$ (its complement in the upper half plane mapping to the other half triangle).  
After interpolating between gridpoints, the solution is analytically continued from $\mathcal{O}_\Delta$ to 
$\mathcal{O}$ by Schwarz reflection, giving the contour plot of figure \ref{contours} (left).

Recall from section \ref{hypertilings} that the octagon vertex and the origin are antipodal points on the Bolza surface.  Placing a 
vortex at the origin breaks most of the symmetry of the Bolza surface, preserving a $C_8$ symmetry both at the vortex position and at its 
antipodal point.  There should thus be a $C_8$ symmetry at the vertex of the octagon when a vortex is placed at the origin, but this is 
not easily seen.  To see it more explicitly, let us shift the vortex to the vertex of the octagon.  There should still be a $C_8$ 
symmetry at the origin.  Recall that there is a double winding of $f$ about the vortex position, which for the vortex at the origin maps 
from the $\{8,8\}$ to the $\{4,8\}$ tessellation.  Angles measured at the vortex position are doubled as illustrated in figures 
\ref{mapf} and \ref{mapf2}.  When the vortex is placed at a vertex, however, it is the angles at the vertices of $\mathcal{O}$ which are 
doubled (from $\pi/4$ to $\pi/2$), while the map is conformal on the interior.  The image is then an octagon in the $\{8,4\}$ 
tessellation.  The absence of ramification points on the interior of the octagon means the degree of the map is $1$ and the 
Riemann-Hurwitz theorem tells us that the area of the image octagon is $A'_{\{8,4\}}=A_{\{8,8\}}-2\pi=2\pi$.  Note that instead of 
\eqref{bbradlow}, the more general formula for the area of the image polygon when $N$ vortices are placed at the vertex is
\begin{equation}%
A'\,=\,A-2\pi N\,=\,2\pi(2g-2-N),
\end{equation}%
which still implies the Bradlow bound \eqref{bradlowagain}.  Dividing the domain and image polygons into triangles as before gives the contours of $|\Phi|^2$ shown 
in figure \ref{contours} (right).
\newpage
\subsection{Magnetic flux}
We now proceed to check consistency of the magnetic flux.  A single vortex has flux $2\pi$, and we 
thus expect to obtain this value when $B$ is integrated over the fundamental octagon.  Using the same 
notation as in \eqref{gaugepotential} the flux through the fundamental octagon $\mathcal{O}$ can be 
computed using Stokes' theorem,\footnote{Recall from section \ref{fproperties} that $|\Phi|$ is invariant under transformations by the 
elements of the Fuchsian group $H$ \eqref{MT}.  It follows from the Bogomolny equations \eqref{bogomolny} that the physical magnetic flux 
density $B/\Omega$ is invariant under these transformations, although neither $B$ nor $\Omega$ is invariant on its own.  The physical 
magnetic flux is thus $\int(B/\Omega)(\Omega\,dxdy)$, where $\Omega\,dxdy$ is the area form on the Poincar\'e disk.  Care must be taken 
to include the total derivative term $d\chi$ which affects the total flux, but does not alter the magnetic field locally.}
\begin{equation}%
\int_\mathcal{O}B\,dxdy\,=\,\int_{\partial\mathcal{O}}a_\mathrm{t}d\ell\,=\,\int_{\partial\mathcal{O}}\left(\tfrac{1}{2}\partial_\mathrm{n}h+\partial_\mathrm{t}\chi\right)d\ell,
\end{equation}%
where $d\ell$ denotes an infinitesimal line element of the boundary of the octagon, $a_\mathrm{t}$ is the tangential component of the gauge potential, 
and $\partial_\mathrm{t}$ and 
$\partial_\mathrm{n}$ are derivatives tangent to and normal to the boundary $\partial\mathcal{O}$.  We note 
from figure \ref{contours} that the midpoints of the edges of the octagon are saddle points of $|\Phi|^2$.  
Combined with the reflection symmetry across the edges of the octagon we conclude that 
$\partial_\mathrm{n}h$ vanishes everywhere on the boundary.

Let us now consider the topological contribution $\int_{\partial\mathcal{O}}\partial_\mathrm{t}\chi\,d\ell$.  On a segment of the 
boundary of $\mathcal{O}$, it evaluates to the difference in $\chi$ at the endpoints, $\Delta\chi$.  From \eqref{generalvortex} we 
have, with $f'=\frac{df}{dz}$,
\begin{equation}%
\chi\,=\,-\frac{\text{i}}{2}\log\left(f'\overline{f'}^{-1}\right).
\end{equation}%
Dividing the octagon into sixteenths as in figure \ref{mapf}, we carry out a rotation $z\mapsto z\text{e}^{\text{i}\pi/4}$ to take us from one vertex of the 
fundamental octagon to the next.  From \eqref{dfdz}, this leads to
$\chi\mapsto\chi+\pi/4$, hence the contribution to the flux from each edge of the octagon is $\pi/4$, and the total 
flux is $2\pi$.  Mapping between opposite edges of the fundamental octagon has the effect 
$\Delta\chi\mapsto\Delta\chi-\pi/2=-\pi/4$, where the change in sign agrees with our traversing of the 
edge in the opposite direction.

It is interesting to compare our results with those for a single vortex placed at the origin of the 
Poincar\'e disk, $\mathbb{H}^2$.  From \eqref{blaschkeproduct}, this vortex has $f(z)=z^2$ and Higgs field \eqref{generalvortex}
\begin{equation}
|\Phi_0|\,=\,\frac{2|z|}{1+|z|^2}.\label{singlevortex}
\end{equation}
Near the origin, $|\Phi_0|=2|z|+\mathcal{O}(|z|^3)$, which should be compared to the result for the regular octagonal tessellation 
\eqref{PhiexactO}, where the leading coefficient of $|\Phi|$ at the vortex is approximately $1.768$.  To get a feel for the change in the 
fields near the origin when a single vortex on $\mathbb{H}^2$ is replaced by an $\{8,8\}$ tessellation of vortices, we integrate the flux of the single vortex over the 
fundamental octagon of figure \ref{figoct}.  As a fraction of the total flux, we find
\begin{equation}%
\frac{1}{2\pi}\int_\mathcal{O}B_0\,dxdy\,=\,\frac{8}{\pi}\,\sqrt{2\sqrt{2}-2}\,\tan^{-1}\left(\sqrt{2\sqrt{2}+2}\right)-2\,\approx\,65.1\%,
\end{equation}%
showing that the lattice of vortices compresses the flux by about $50\%$.

\section{Vortices on higher genus surfaces}\label{highergenus}
The vortex equations on surfaces with $g>2$ can be studied using similar arguments to those given in 
the preceding sections.  The obvious extension of our construction is to a vortex 
at the origin of the Poincar\'e disk in the hyperbolic tessellation with Schl\"afli symbol $\{4g,4g\}$.  Quotienting by the Fuchsian group 
of the tessellation gives a vortex on a compact surface of genus $g$ and area $4\pi(g-1)$.  The fundamental $4g$-gon is divided into $8g$ congruent triangles 
with angles $\left(\tfrac{\pi}{4g},\tfrac{\pi}{2},\tfrac{\pi}{4g}\right)$.  The map $f$ is a two-to-one map from the $4g$-gon 
to a $2g$-gon in the $\{2g,4g\}$ tessellation, and the image of the fundamental triangle is a triangle with angles $\left(\tfrac{\pi}{2g},\tfrac{\pi}{2},\tfrac{\pi}{4g}\right)$.  
Applying the construction described in section \ref{mappingbetweenpolygons} gives the maximum value of $|\Phi|$, measured 
at the vertex of the $4g$-gon,
\begin{equation}
|\Phi|_{\text{max}}\,=\,\frac{\Gamma\!\left(\tfrac{1}{4g}\right)\Gamma\!\left(\tfrac{3}{2g}\right)}{2\,\Gamma\!\left(\tfrac{3}{4g}\right)\Gamma\!\left(\tfrac{1}{g}\right)}\,\sqrt{2-\text{sec}\left(\tfrac{\pi}{2g}\right)},\label{hypergenus}
\end{equation}
and the coefficient of the leading term of $|\Phi|$ near the vortex at the origin,
\begin{equation}
|\Phi|_{\text{origin}}\,=\,\frac{\Gamma^3\!\left(\tfrac{1}{4g}\right)\Gamma\!\left(\tfrac{1}{2g}\right)\Gamma\!\left(\tfrac{3}{2g}\right)}{8\pi g\,\Gamma\!\left(\tfrac{3}{4g}\right)\Gamma^2\!\left(\tfrac{1}{g}\right)}\sqrt{\frac{\text{cot}^2\left(\tfrac{\pi}{4g}\right)-3}{\text{cot}^2\left(\tfrac{\pi}{4g}\right)-1}}\,\,|z|+\mathcal{O}\left(|z|^3\right).\label{hypergenusorigin}
\end{equation}
As the genus is increased, the area available to a vortex grows linearly with $g$.  The resulting increase in 
separation between adjacent vortices in the tessellation suggests that for large $g$, $|\Phi|$ should approach the solution $|\Phi_0|$ 
for a single vortex in $\mathbb{H}^2$ \eqref{singlevortex}, which has $|\Phi_0|_{\text{max}}=1$ and $|\Phi_0|_{\text{vortex}}=2|z|+\mathcal{O}(|z|^3)$.  Evaluating 
\eqref{hypergenus} and the leading coefficient of \eqref{hypergenusorigin} for some specific values of $g$, we do indeed 
approach this limit:

\begin{center}
\begin{tabular}{r|ccccc}
&$g=2$&$g=3$&$g=4$&$\cdots$&$g\to\infty$\\\hline\\[-10pt]
$|\Phi|_{\text{max}}^{\phantom{\text{origin}}}$&0.841&0.965&0.986& &1\\[5pt]
$|z|^{-1}|\Phi|_{\text{origin}}$&1.768&1.955&1.984& &2
\end{tabular}
\end{center}

The principal interest of higher genus surfaces is that the Bradlow bound (\ref{bradlow}, \ref{bradlowagain}) allows them to hold more than 
one vortex.  For example, for $g=3$ the regular $\{12,12\}$ tessellation with opposite edges identified allows 
one, two or three vortices placed at the origin or a vertex of the dodecagon.  The Higgs field can again be computed by dividing the 
dodecagon into congruent triangles.  Placing $N_\text{o}$ vortices at the origin and $N_\text{v}$ at the vertex of the dodecagon requires 
a map $f$ with ramification of order $N_\text{o}+1$ at the origin and $N_\text{v}+1$ at the vertex.  Thus the image tessellation is of 
type
\begin{equation*}
\left\{\frac{12}{N_\text{o}+1},\frac{12}{N_\text{v}+1}\right\},
\end{equation*}
whose fundamental polygon has an area $A'$ given by either \eqref{areaformula} or the Riemann-Hurwitz theorem:
\begin{equation}%
A'\,=\,2\pi\,\frac{4-N_\text{o}-N_\text{v}}{N_\text{o}+1}.
\end{equation}%
The Bradlow bound requires that $N_{\text o}+N_{\text v}<4$.  This ensures that $N_{\text o}+1$ and $N_{\text v}+1$ both divide $12$ and 
that the image tessellation is hyperbolic.  The Euclidean tessellations of types $\{6,3\}$ and $\{4,4\}$ do not occur, because they would 
both require $N_{\text o}+N_{\text v}=4$.  It is easy to check using the expansion of the Schwarz triangle function \eqref{szero} that 
$|\Phi|_\text{origin}$ behaves like $|z|^{N_\text{o}}$, with angular dependence at order $|z|^{N_\text{o}+4g}$.  Note that on more 
general $\{p,q\}$ tessellations, integer winding of $f$ imposes divisibility constraints on $N_\text{o}$ and $N_\text{v}$.

We remark that the surface described by the $\{12,12\}$ tessellation with opposite edges identified is not the most symmetric genus $3$ 
surface, which is the so-called Klein surface that can be represented by a $\{14,7\}$ tessellation.  However, our construction only 
allows us to place a single vortex on this surface, necessarily at the origin of the $14$-gon (as $p\neq q$ the centres and vertices of 
the polygon are not equivalent).  The image of the $14$-gon is a heptagon in the $\{7,7\}$ tessellation, and we obtain a vortex with 
$C_7$ symmetry with $|\Phi|_{\text{max}}\approx0.952$ and $|\Phi|_{\text{origin}}\approx1.961|z|+\dots$.
\newpage
\subsection{A further quotient}
\begin{figure}
\begin{minipage}{0.485\linewidth}
\centering
\vspace{-0.3cm}
\includegraphics[width=0.75\linewidth]{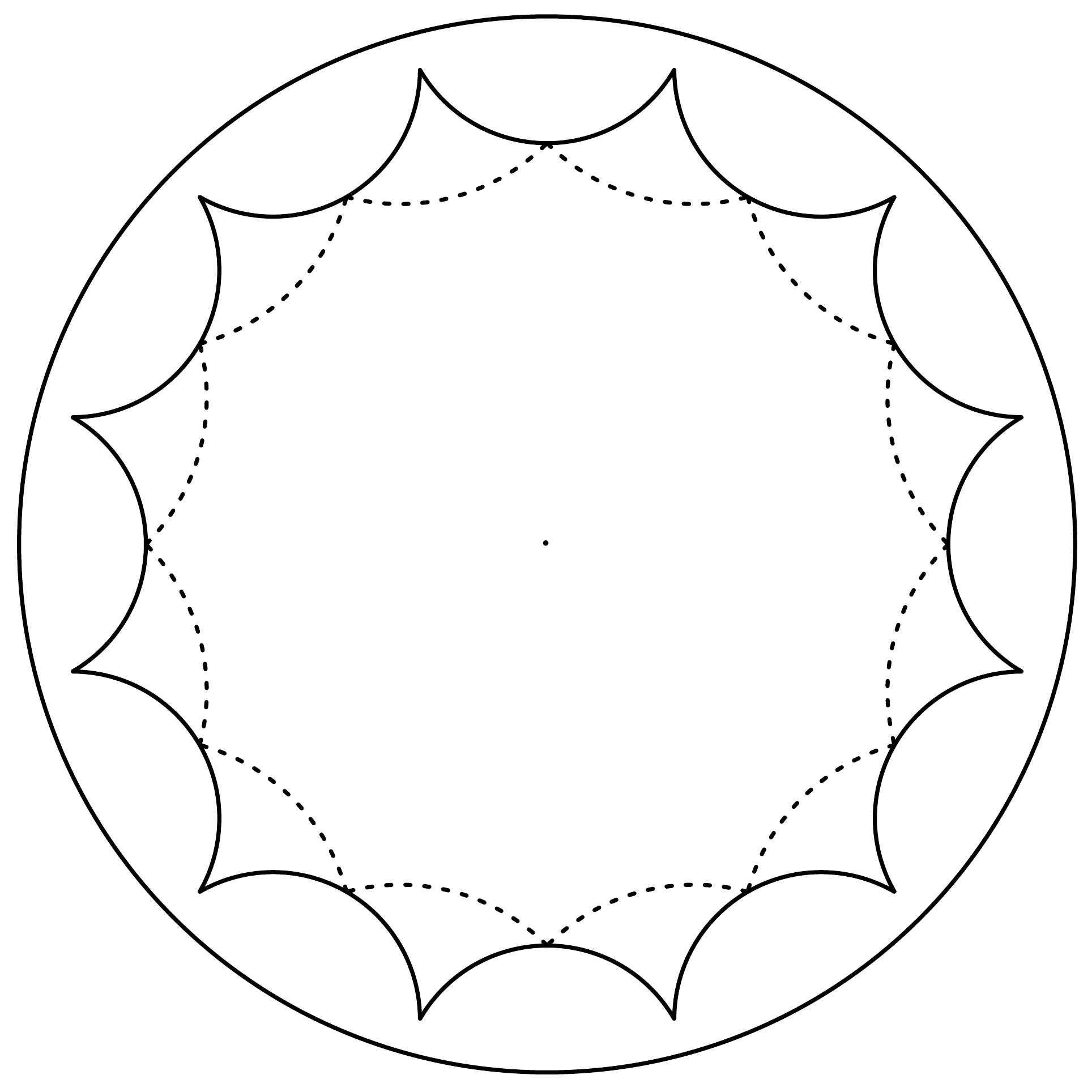}
\end{minipage}
\begin{minipage}{0.485\linewidth}
\centering
\includegraphics[width=0.75\linewidth]{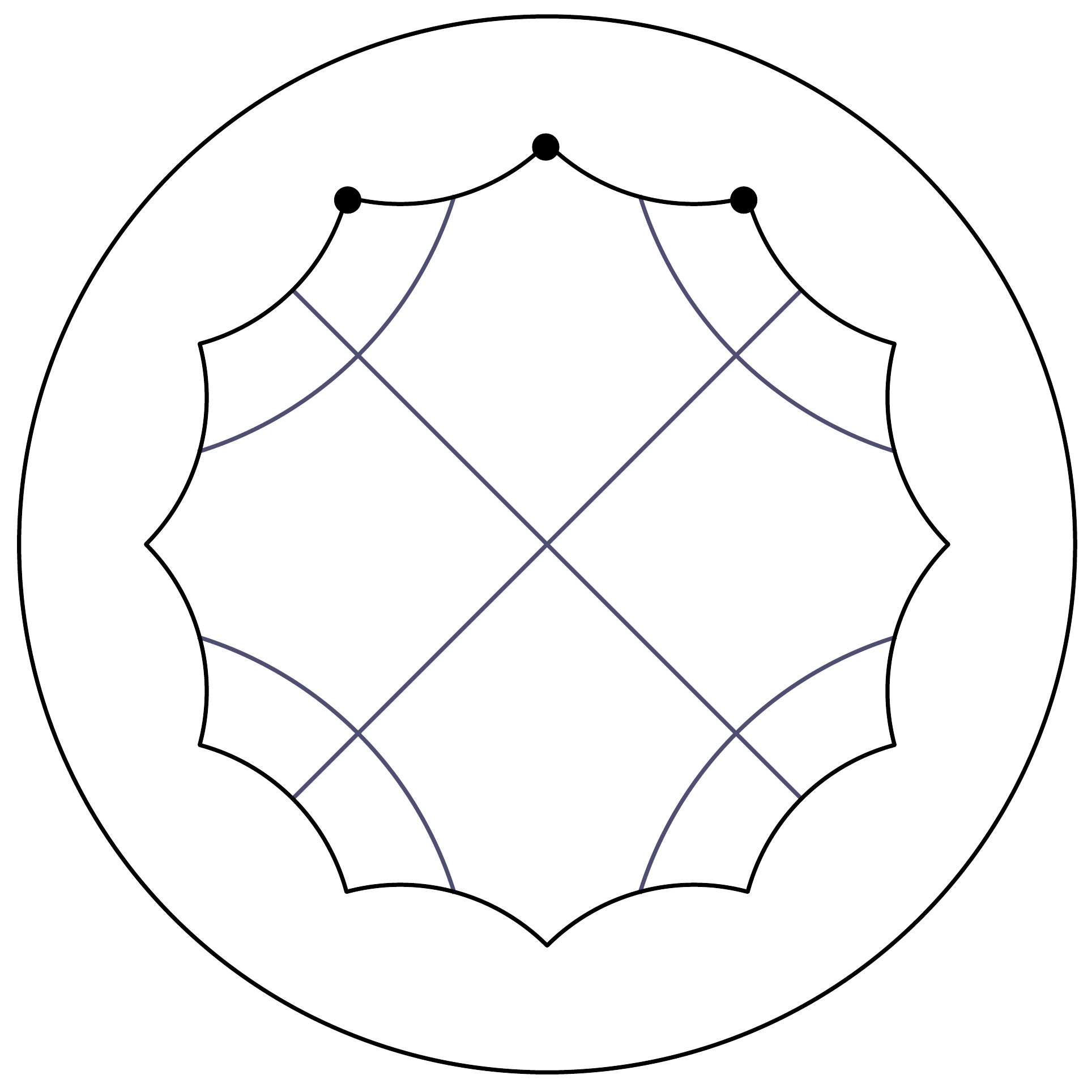}
\end{minipage}
\caption{Left:~the $\{12,12\}$ tessellation with opposite edges identified (solid line) can be cut along the 
dotted line to give a dodecagon of half the area in the $\{12,4\}$ tessellation.  Identifying edges as 
shown in the right hand diagram yields a smooth compact genus $2$ surface.  Placing a vortex at the 
origin of the $\{12,12\}$ cell and another at the vertex is therefore equivalent to a single vortex 
at the origin of a $\{12,4\}$ cell.  Three distinct vertices are indicated by dots.}\label{dodecagon}
\end{figure}

A case of special interest is the $\{12,12\}$ tessellation with one vortex placed at the origin of the 
dodecagon and one at the vertex.  The image tessellation is of type $\{6,6\}$.  The leading behaviour of 
the Higgs field near each vortex is
\begin{equation}
|\Phi|_{\text{origin}}\,=\,\frac{2-\sqrt{3}}{36\sqrt{2}}\,\frac{\Gamma^4(\tfrac{1}{12})}{\Gamma^4(\tfrac{1}{3})}\,|z|+\mathcal{O}(|z|^3)\,\approx\,1.787|z|+\mathcal{O}(|z|^3).\label{dodechiggs}
\end{equation}
We find the same leading coefficient when placing a single vortex at the origin of a dodecagon in 
the $\{12,4\}$ tessellation.  This result is due to the existence of a fixed-point-free $\mathbb{Z}_2$ 
action on the genus $3$ surface which identifies the origin and vertex of the $\{12,12\}$ tessellation.  
Quotienting by this group gives a smooth compact genus $2$ surface with one vortex on it.  This procedure is only possible 
when the vortex locations are compatible with the $\mathbb{Z}_2$ symmetry, and the quotient surface 
contains half the number of vortices.  Figure \ref{dodecagon} illustrates the relation between the 
two tessellations.

This observation can be understood by explicitly looking for a free $\mathbb{Z}_2$ action, an involution $\mathcal{R}$ on the 
hyperelliptic curve.  Let us generalise our discussion to a genus $g$ surface represented by a regular $\{4g,4g\}$ tessellation with 
opposite edges identified, with a vortex at the centre and another at the vertex.  As a Riemann surface, this genus $g$ surface is a 
double cover of $\mathbb{CP}^1$ with $2(g+1)$ ramification points.  The ramification points at the north and south poles correspond to 
the centre and vertex of the polygon, and there are $2g$ equally spaced ramification points on the equator.  Generalising what was 
discussed in section \ref{hypertilings} for the $g=2$ Bolza surface, the cyclic $C_{4g}$ symmetry has fixed points at the polygon centre 
and at the vertex, while there are $g$ $C_4$ symmetries with pairs of fixed points on the equator.  The symmetry group of the surface is 
a double group of the dihedral group $D_{2g}$, and the hyperelliptic curve representing this setup is\footnote{In the special case of 
the Bolza surface ($g=2$), the symmetry is enhanced to $\text{GL}(2,3)$, a double cover of the cubic group.}
\begin{equation}%
y^2\,=\,x(x^{2g}+1),\qquad\qquad(x,y)\,\in\,\left(\mathbb{C}^\ast\right)^2.
\end{equation}%
The desired involution must map between the origin of the polygon and the vertex (i.e.~between the poles of the sphere $x=0$ and 
$x=\infty$) and have no fixed points.  This is only possible for $g$ odd (the case $g$ even has fixed 
points at $(x,y)=(-1,\pm\sqrt{2}\,\text{i})$), in which case it is achieved 
by the involution
\begin{equation}%
\mathcal{R}:(x,y)\,\mapsto\,\left(x^{-1},-y\,x^{-g-1}\right).
\end{equation}%
The quotient of the branched cover of $\mathbb{CP}^1$ by this involution is a double covering of the northern hemisphere with a 
particular identification of the segments of the equator between ramification points.  In the polygon picture, there are $4g$ edges to pair 
up.  Consistency with the transformation $x\mapsto x^{-1}$ gives two possible pairings for each edge, due to the choice of branch.  We 
also require the quotient surface to be smooth, which will be the case if there is no angle deficit at the identified vertices.  The 
total internal angle of the quotient $\{4g,4\}$ polygon is $2\pi g$, hence there must be $g$ distinct vertices when the edges are 
paired.  This imposes a further constraint on our choice of edge identifications.  There is also the possibility that edge pairs can be 
twisted, although it is not clear whether our construction, which relies on $f(z)$ extending from a fundamental triangle to all of 
$\mathbb{H}^2$ by reflections, works in the twisted case.  Once the correct identifications have been made, the resulting surface has 
half the area of the original surface and genus $\tfrac{1}{2}(g+1)$.

The identification described in the preceding paragraphs is illustrated for the $g=3$ case in figure 
\ref{dodecagon}.  The genus $2$ surface in the $\{12,4\}$ tessellation has different Fenchel-Nielsen 
parameters from the $\{8,8\}$ tessellation, and they therefore represent different surfaces.  An 
immediate consequence of this is the different leading behaviour of the Higgs field, as is seen by 
contrasting \eqref{PhiexactO} and \eqref{dodechiggs}.  Because a $2$-holed surface has no freely 
acting rotations by $\pi$, the genus cannot be reduced further.
\section{Conclusions and outlook}\label{conclusions}
In this paper we have shown how to construct Abelian-Higgs vortices on compact hyperbolic surfaces which can be represented by 
tessellations of the hyperbolic plane by regular polygons.  Placing vortices on the surface breaks most of the symmetry, and our method 
requires the vortex positions to be chosen in such a way that there is a preserved symmetry allowing the surface to be covered by 
congruent triangles sharing vertices at the vortex positions.  In particular, multivortex solutions can be constructed with vortices 
located at antipodal points on the surface, corresponding to the origin and vertex of the polygonal tessellation.  The Higgs field is 
computed by identifying a holomorphic function on the fundamental polygon which is conformal everywhere except at the vortex positions, 
where there is a ramification point of order one greater than the number of vortices at that location.  This function can be expressed 
implicitly as the composition of two Schwarz triangle functions, allowing numerical evaluation of the Higgs field at any given point on 
the surface.

There are several immediate generalisations of this procedure which it would be interesting to investigate further:
\begin{itemize}
 \item The tessellation can be deformed such that the angles are fixed but the edge lengths change (see figure \ref{def}).  In this case we 
 require a map from a quadrant of the fundamental octagon to half a kite.  The construction is analogous to that of the Schwarz triangle 
 map described in section \ref{solutions}, with the hypergeometric differential equation \eqref{riemannde} replaced by the Heun 
 differential equation, which has four regular singular points.  The difficulty with this construction is the appearance  of an 
 {\it accessory parameter} which characterises the deformation, and which requires numerical computation.  A different deformation of the 
 regular octagon was studied in \cite{Naz}.
 \begin{figure}
\begin{minipage}{0.485\linewidth}
\centering
\vspace{-0.3cm}
\includegraphics[width=0.75\linewidth]{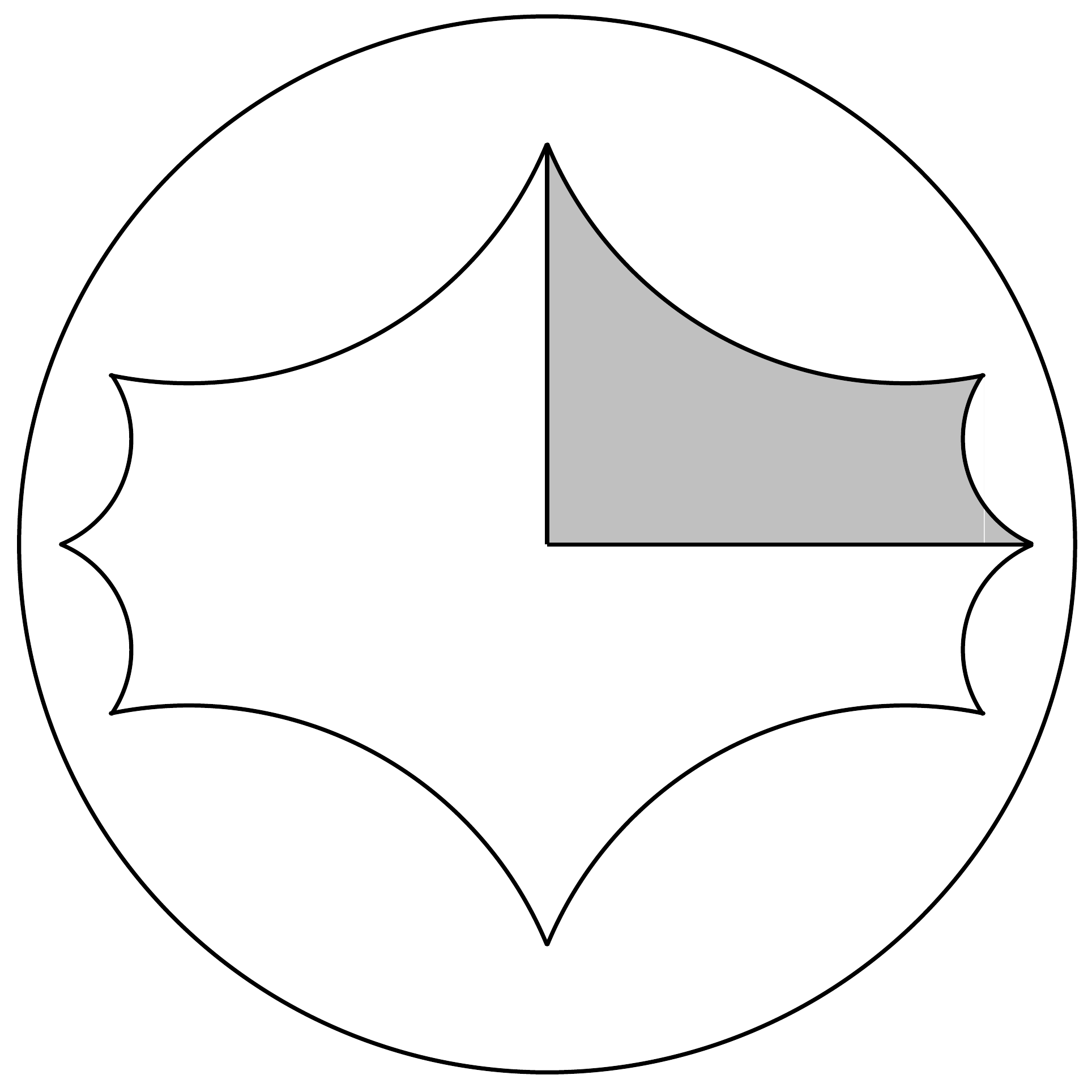}
\end{minipage}
\begin{minipage}{0.485\linewidth}
\centering
\includegraphics[width=0.75\linewidth]{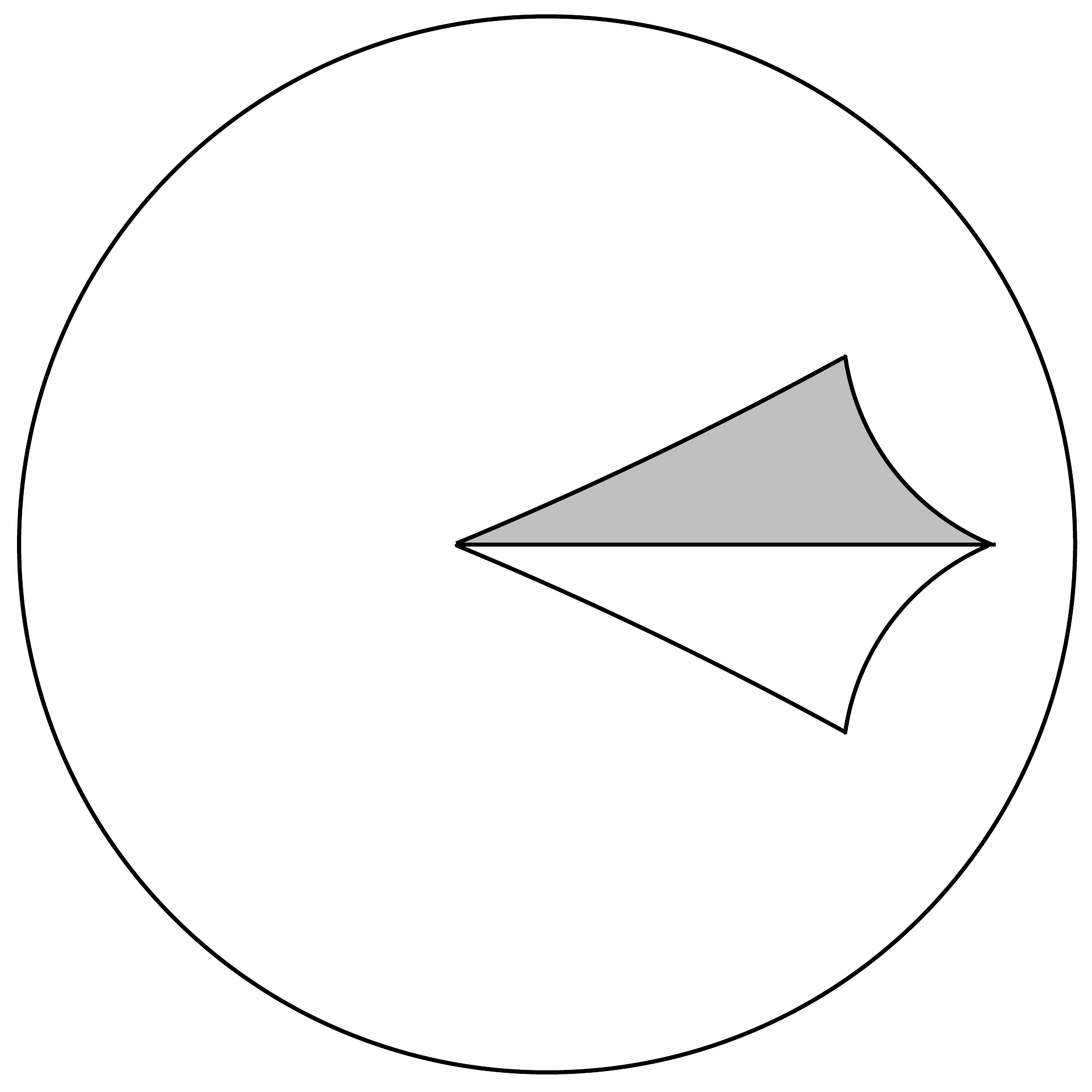}
\end{minipage}
\caption{An example of a deformed $\{8,8\}$ octagon with angles of $\pi/4$ which can be split into four quadrilaterals, each of which 
maps to half of a deformed image quadrilateral.}\label{def}
\end{figure}
 \item The construction described in this paper does not allow the vortex to be placed at an arbitrary position on the surface.  The 
 Riemann map from a hyperbolic polygon to the upper half plane is analytically continued to cover the Poincar\'e disk by reflection in 
 the edges of the polygon.  If the vortex is not positioned symmetrically then the fields cannot be constructed by maps between 
 regular polygons.  As it stands, our construction only enables us to compute the fields at isolated points in the moduli space.
 \item A final outstanding question is how to place an arbitrary number $N$ of vortices (with $N<2g-2$) at the origin of a symmetric 
 tessellation  describing a genus $g$ surface.  An $(N+1)$-fold map from the fundamental polygon to a smaller polygon is only possible if 
 $N+1$ divides the number of edges of the initial polygon.  For example, we can place a single vortex on the Klein surface (with a map 
 from a $14$-gon to a heptagon), but not two or three vortices.
\end{itemize}

\section*{Acknowledgements}
This work was supported by the UK Science and Technology Facilities Council under grant number ST/J000434/1.


{\small\begin{thebibliography}{9}
\raggedright
\bibitem{JT80}A.~Jaffe, C.~Taubes, \emph{Vortices and Monopoles}, Birkh\"auser (1980)
\bibitem{Bap14}J.~M.~Baptista, \emph{Vortices as degenerate metrics}, Lett.~Math.~Phys.~{\bf 104} (2014) 731, \href{http://arxiv.org/abs/1212.3561}{\texttt{arXiv:1212.3561 [hep-th]}}
\bibitem{Wit77}E.~Witten, \emph{Some exact multipseudoparticle solutions of classical Yang-Mills theory}, Phys.~Rev.~Lett.~{\bf 38} (1977) 121
\bibitem{ET67}U.~Essmann, H.~Tr\"auble, \emph{The direct observation of individual flux lines in type II superconductors}, Phys.~Lett.~{\bf A 24} (1967) 526
\bibitem{MR10}N.~S.~Manton, N.~A.~Rink, \emph{Vortices on hyperbolic surfaces}, J.~Phys.~A {\bf 43} (2010) 434024, \href{http://arxiv.org/abs/0912.2058}{\texttt{arXiv:0912.2058 [hep-th]}}
\bibitem{FP87}W.~J.~Floyd, S.~P.~Plotnick, \emph{Growth functions on Fuchsian groups and the Euler characteristic}, Invent.~Math.~{\bf 88} (1987) 1
\bibitem{Naz}A.~V.~Nazarenko, \emph{Two-parametric hyperbolic octagons and reduced Teichm\"uller space in genus two}, \href{http://arxiv.org/abs/1301.5446}{\texttt{arXiv:1301.5446 [math-ph]}}
\bibitem{Neh52}Z.~Nehari, \emph{Conformal Mapping}, McGraw-Hill (1952)
\bibitem{HM03}M.~Harmer, G.~Martin, \emph{Conformal mappings from the upper half plane to fundamental domains on the hyperbolic plane}, Department of Maths Report Series 499, University of Auckland, N.Z.~(2003)


\end{thebibliography}}
\end{document}